\documentclass[a4paper,reqno]{amsart}
\usepackage{amsmath,amssymb,mathrsfs}
\usepackage{txfonts,bm,pifont}
\usepackage{cite}
\usepackage[dvipdfm,bookmarks=true,bookmarksnumbered=true,colorlinks=true]{hyperref}
\usepackage{comment}
\usepackage[dvipdfmx]{graphicx}
\PassOptionsToPackage{svgnames}{xcolor}
\usepackage{tikz}
\usetikzlibrary{calc}
\usetikzlibrary{decorations.markings,decorations.pathmorphing}

\newtheorem{theorem}{Theorem}[section]

\newtheorem{remark}[theorem]{Remark}

\numberwithin{equation}{section}
\newcommand{\wutilde}[1]{\vrule depth 0pt width 0pt%
{\raise0.8pt\hbox{$\smash{{\mathop{#1} \limits_{\displaystyle\widetilde{}}}}$}}}

\newcommand{\hT}{\hat{T}}
\newcommand{\hR}{\hat{R}}
\newcommand{\al}{\alpha}
\newcommand{\be}{\beta}
\newcommand{\de}{\delta}
\newcommand{\ga}{\gamma}
\newcommand{\hal}{\hat{\al}}
\newcommand{\hbe}{\hat{\be}}

\newcommand{\hga}{\hat{\ga}}

\newcommand{\la}{\lambda}
\newcommand{\ep}{\bm{\epsilon}}
\newcommand{\PDE}{P$\Delta$E}
\newcommand{\bbZ}{\mathbb{Z}}

\newcommand{\bbC}{\mathbb{C}}
\newcommand{\calR}{{\mathcal R}}
\newcommand{\calr}{{\mathcal r}}
\newcommand{\ii}{{\rm i}}
\newcommand{\bml}{{\bm l}}
\newcommand{\IIIT}{\hat{T}_{\rm III}}
\newcommand{\IVT}{\hat{T}_{\rm IV}}
\newcommand{\SIIIT}{\hat{T}_{\rm SIII}}
\newcommand{\spT}{\hat{T}_{\rm SP}}
\newcommand{\Set}[2]{\left\{#1\right\}_{#2}}
\newcommand{\set}[2]{\left\{\left. #1 ~\right|~ #2 \right\}}
\makeatletter
\long\def\@makecaption#1#2{
 \vskip 10pt
 \setbox\@tempboxa\hbox{#1. #2}
 \ifdim \wd\@tempboxa >\hsize #1. #2\par \else \hbox
to\hsize{\hfil\box\@tempboxa\hfil}
 \fi}
\begin{document}
\title{Lax pairs of discrete Painlev\'e equations: $(A_2+A_1)^{(1)}$ case}
\author{Nalini Joshi}
\author{Nobutaka Nakazono}
\address{School of Mathematics and Statistics, The University of Sydney, New South Wales 2006, Australia.}
\email{nobua.n1222@gmail.com}
\begin{abstract}
In this paper, we provide a comprehensive method for constructing Lax pairs of discrete Painlev\'e equations by using a reduced hypercube structure.
In particular, we consider the $A_5^{(1)}$-surface $q$-Painlev\'e system 
which has the affine Weyl group symmetry of type $(A_2+A_1)^{(1)}$. Two new Lax pairs are found.  
\end{abstract}

\subjclass[2010]{
33E17, 
37K10, 
39A13, 
39A14}
\keywords{ 
Discrete Painlev\'e equation; 
ABS equation; 
Lax pair; 
$\tau$ function;
affine Weyl group}
\maketitle
\setcounter{tocdepth}{1}

\section{Introduction}
The purpose of this paper is to provide a comprehensive method for constructing Lax pairs of discrete Painlev\'e equations by using a reduced hypercube structure arising from $\omega$-lattices,
composed from the $\tau$ functions of discrete Painlev\'e equations (see \cite{JNS2015:MR3403054,JNS:paper5} for details).
The term ``reduced hypercube structure" is used to describe periodic reduction of lattices obtained from multi-dimensionally consistent hypercubes (see \S \ref{subsection:background}).
As an example, we demonstrate the constructions of Lax pairs of the $q$-Painlev\'e equations \eqref{eqns:qpainleve_intro}.

Our previous work \cite{JNS2015:MR3403054,JNS2014:MR3291391} brought together a lattice in higher dimensions with $\tau$-function theory, and showed how a geometric reduction provided an $\omega$-lattice leading to $A_5^{(1)}$-surface $q$-Painlev\'e equations. 
Here, we show how this perspective enables us to systematically construct the Lax pairs for any discrete Painlev\'e equations on the $A_5^{(1)}$-surface. 
Note that in \cite{SakaiH2001:MR1882403} Sakai classified the discrete Painlev\'e equations into 22 surface types 
according to the configuration of the base points (i.e. points where the system is ill defined because it approaches $0/0$) as the following: 
\begin{center}
\begin{tabular}{|l|l|}
\hline
Discrete type&Type of surface\\
\hline
Elliptic&$A_0^{(1)}$\rule[-.5em]{0em}{1.6em}\\
\hline
Multiplicative&$A_0^{(1)}$, \dots, $A_8^{(1)}$, $A_7^{(1)'}$\rule[-.5em]{0em}{1.6em}\\
\hline
Additive&$A_0^{(1)}$, $A_1^{(1)}$, $A_2^{(1)}$, $D_4^{(1)}$, \dots, $D_8^{(1)}$, $E_6^{(1)}$, $E_7^{(1)}$, $E_8^{(1)}$\rule[-.5em]{0em}{1.6em}\\
\hline
\end{tabular}
\end{center}
There are possibly infinitely many discrete Painlev\'e equations on each surface. 
In this paper, we use the collective term ``$A_5^{(1)}$-surface $q$-Painlev\'e equations"
for discrete Painlev\'e equations on the multiplicative $A_5^{(1)}$-surface. 

Our Lax pairs for the 
$q$-Painlev\'e IV equation \eqref{eqn:qp4_intro} and $q$-Painlev\'e III equation \eqref{eqn:qp3_intro} are new, 
while the one for the $q$-Painlev\'e II equation \eqref{eqn:qp2_intro}
coincides with that provided in \cite{HHJN2007:MR2303490}. The spectral part of these Lax pairs is more regular than those from other known Lax pairs. More specifically, it satisfies Carmichael's hypotheses \cite{CarmichaelRD1912:MR1506145} for existence of solutions around singular points at the origin and infinity. These properties allowed Carmichael to prove existence of solutions in a way that is  considered to be the $q$-analogue of Fuchsian theory. 

A well-known example which also satisfies Carmichael's conditions is the Lax pair for the $q$-Painlev\'e VI equation ($q$-P($A_3^{(1)}$)), constructed by Jimbo and Sakai  \cite{JS1996:MR1403067}. This was seen as a natural analogue of the Fuchsian nature of the Lax pair for the sixth Painlev\'e equation \cite{FuchsR1905:quelques}. However, the Lax pairs of other $q$-Painlev\'e equations \cite{MurataM2009:MR2485835}, obtained by degeneration of the Lax pair for $q$-P($A_3^{(1)}$), no longer satisfy Carmichael's hypotheses: the coefficient matrix of either the highest degree or the lowest degree term in the spectral variable is singular. Our results suggest that contrary to expectation, such $q$-Painlev\'e equations may still have regular Lax pairs. We provide three examples to support this suggestion.

It is not surprising to have multiple Lax pairs for discrete Painlev\'e equations as multiple ones are known for each of the continuous Painlev\'e equations \cite{JKT2009:MR2525382,JKT2007:MR2362885,DM2000:MR1767271}. 
\subsection{$A_5^{(1)}$-surface $q$-Painlev\'e equations}
The $q$-difference equations we study are
\begin{subequations}\label{eqns:qpainleve_intro}
\begin{align}
 \text{$q$-P$_{\rm IV}$ : }
 &\cfrac{\overline{f}}{abg}=\cfrac{1+c h(a f+1)}{1+a f(b g+1)},\quad
 \cfrac{\overline{g}}{bch}=\cfrac{1+a f(b g+1)}{1+b g(c h+1)},\quad
 \cfrac{\overline{h}}{caf}=\cfrac{1+b g(c h+1)}{1+c h(a f+1)},
 \label{eqn:qp4_intro}\\
 \text{$q$-P$_{\rm III}$ : }
 &\overline{g}g=\cfrac{a(1+t f)}{f(t+f)},\quad
 \overline{f}f=\cfrac{a(1+b t \overline{g})}{\overline{g}(bt+\overline{g})},
 \label{eqn:qp3_intro}\\
 \text{$q$-P$_{\rm II}$ : }
 &\widetilde{f}\wutilde{f}=\cfrac{a(1+tf)}{f(t+f)},
 \label{eqn:qp2_intro}
\end{align}
\end{subequations}
where $t,a,b,c,q,p\in\bbC^\ast$ and
\begin{subequations}
\begin{align}
 &f=f(t),\quad
 g=g(t),\quad
 h=h(t),\quad
 \overline{f}=f(q t),\quad
 \overline{g}=g(q t),\quad
 \overline{h}=h(q t),\\
 &\widetilde{f}=f(p t),\quad 
 \wutilde{f}=f(p^{-1}t).
\end{align}
\end{subequations}
In the case of $q$-P$_{\rm IV}$, we have the following conditions: 
\begin{equation}
 fgh=t^2,\quad abc=q.
\end{equation}
We note that $q$-P$_{\rm IV}$, $q$-P$_{\rm III}$ and $q$-P$_{\rm II}$ are known as 
a $q$-discrete analogue of the Painlev\'e IV equation \cite{KNY2001:MR1876614},
that of the Painlev\'e III equation \cite{KTGR2000:MR1789477,SakaiH2001:MR1882403} and
of the Painlev\'e II equation \cite{RG1996:MR1399286},
respectively.

\begin{remark}
It is known that
$q$-{\rm P}$_{\rm III}$ \eqref{eqn:qp3_intro} can be reduced to 
$q$-{\rm P}$_{\rm II}$ \eqref{eqn:qp2_intro} by projective reduction \cite{KNT2011:MR2773334}:
\begin{equation}
 b=p,\quad q=p^2,\quad g=\wutilde{f}.
\end{equation}
In this sense, $q$-{\rm P}$_{\rm II}$ is often described as the scalar form of $q$-{\rm P}$_{\rm III}$.
Although the projective reduction is a simple specialization of the parameters at the level of the equation, 
the resulting equations have different type of hypergeometric solutions (see \cite{KN2015:MR3340349,KNT2011:MR2773334} and references therein).
In this paper, we also show that they have different Lax pairs 
but share the same spectral linear problem (see \S \ref{subsection:main_result}).
\end{remark}

\subsection{Main results}
\label{subsection:main_result}
Our main result shows that Equations \eqref{eqns:qpainleve_intro} share one spectral linear problem, 
which takes a factorized form
\begin{equation}\label{eqn:intro_linear}
 \phi(qx)=A\phi(x),
\end{equation}
where
\begin{equation}\label{eqn:linear_martrix_A}
 A
 =\begin{pmatrix}
 - \cfrac{\ii q\la}{f_2}\,x&1\\
 -1&- \cfrac{\ii qf_2}{\la}\,x
 \end{pmatrix}
 \begin{pmatrix}
 - \cfrac{\ii\la a_0a_2}{f_0}\,x&1\\
 -1&- \cfrac{\ii a_0a_2 f_0}{\la}\,x
 \end{pmatrix}
 \begin{pmatrix}
 - \cfrac{\ii\la a_0}{f_1}\,x&1\\
 -1&- \cfrac{\ii a_0 f_1}{\la}\,x
 \end{pmatrix}.
\end{equation}
Here, the non-zero complex parameters $a_i$, $i=0,1,2$, $\la$ and $q$ and the variables $f_i$, $i=0,1,2$,  satisfy
\begin{equation}
 a_0a_1a_2=q,\quad
 f_0f_1f_2=\la^2.
\end{equation}
However, the deformation problem in the Lax pairs differs for different cases in Equations \eqref{eqns:qpainleve_intro}.
Let $\IVT$, $\IIIT$ and $\SIIIT$ be deformation operators
whose actions on the parameters $a_i$, $i=0,\dots,2$, $\la$ and $q$ are given by
\begin{subequations}\label{eqns:intro_action_para}
\begin{align}
 &\IVT:(a_0,a_1,a_2,\la,q)\mapsto (a_0,a_1,a_2,q\la,q),\\
 &\IIIT:(a_0,a_1,a_2,\la,q)\mapsto (qa_0,q^{-1}a_1,a_2,\la,q),\\
 &\SIIIT:(a_0,a_1,a_2,\la,q)\mapsto (a_0a_2,q^{-1}a_1a_2,q{a_2}^{-1},\la,q),
\end{align}
\end{subequations}
while those on the spectral parameter $x$ and the wave function $\phi=\phi(x)$ are given by
\begin{subequations}\label{eqns:intro_deformations}
\begin{align}
 &\IVT(x)=\IIIT(x)=\SIIIT(x)=x,\\
 &\IVT(\phi)=B_{\rm IV}\,\phi,\quad
 \IIIT(\phi)=B_{\rm III}\,\phi,\quad
 \SIIIT(\phi)=B_{\rm SIII}\,\phi,
\end{align}
\end{subequations}
where 
\begin{subequations}\label{eqns:linear_martrix_B}
\begin{align}
 B_{\rm IV}
 &=\begin{pmatrix}
 \cfrac{\ii(q\la^2-1)f_2}{\la(1+a_1(1+a_2 f_2)f_1)}\,x&-1\\
 1&0
 \end{pmatrix},\\
 B_{\rm III}
 &=\begin{pmatrix}
 - \cfrac{\ii\la a_0a_2}{f_0}\,x&1\\
 -1&- \cfrac{\ii a_0a_2 f_0}{\la}\,x
 \end{pmatrix}
 \begin{pmatrix}
 - \cfrac{\ii\la a_0}{f_1}\,x&1\\
 -1&- \cfrac{\ii a_0 f_1}{\la}\,x
 \end{pmatrix},\\
 B_{\rm SIII}
 &=\begin{pmatrix}
 - \cfrac{\ii\la a_0}{f_1}\,x&1\\
 -1&- \cfrac{\ii a_0 f_1}{\la}\,x
 \end{pmatrix}.
\end{align}
\end{subequations}
The subscripts ${\rm IV}$, ${\rm III}$, ${\rm SIII}$ label the deformation operators and matrices corresponding to 
$q$-P$_{\rm IV}$ \eqref{eqn:qp4_intro}, $q$-P$_{\rm III}$ \eqref{eqn:qp3_intro},
the scalar form of $q$-{\rm P}$_{\rm III}$ \eqref{eqn:qp2_intro}, respectively.
Equations \eqref{eqns:intro_action_para} and \eqref{eqns:intro_deformations} provide us with the deformation of the spectral problem.

\begin{theorem}\label{maintheorem_Lax}
The compatibility conditions of the linear equation \eqref{eqn:intro_linear} with the operators 
$\IVT$, $\IIIT$ and $\SIIIT$$:$
\begin{equation}
 \IVT(A)B_{\rm IV}=B_{\rm IV}(qx)A,\quad
 \IIIT(A)B_{\rm III}=B_{\rm III}(qx)A,\quad
 \SIIIT(A)B_{\rm SIII}=B_{\rm SIII}(qx)A,
\end{equation}
are equivalent to
{\allowdisplaybreaks
\begin{subequations}\label{eqns:intro_compatibility_conditions}
\begin{align}
 &\begin{cases}
 ~\cfrac{\IVT(f_0)}{a_0a_1 f_1}=\cfrac{1+a_2 f_2(a_0 f_0+1)}{1+a_0 f_0(a_1 f_1+1)},\\
 ~\cfrac{\IVT(f_1)}{a_1a_2 f_2}=\cfrac{1+a_0 f_0(a_1 f_1+1)}{1+a_1 f_1(a_2 f_2+1)},\\
 ~\cfrac{\IVT(f_2)}{a_2a_0 f_0}=\cfrac{1+a_1 f_1(a_2 f_2+1)}{1+a_2 f_2(a_0 f_0+1)},
 \end{cases}\label{eqn:intro_T4action_f}\\
 &\IIIT(f_1)f_1=\cfrac{\la^2(1+a_0f_0)}{f_0(a_0+f_0)},\quad
 \IIIT(f_0)\IIIT(f_1)=\cfrac{\la^2(1+a_0a_2\IIIT(f_1))}{f_0(a_0a_2+\IIIT(f_1))},
 \label{eqn:intro_T3action_f}\\
 &\SIIIT(f_1)=f_0,\quad
 \SIIIT(f_0)f_1=\cfrac{\la^2(1+a_0f_0)}{f_0(a_0+f_0)},
 \label{eqn:intro_TS3action_f}
\end{align}
\end{subequations}
respectively.
}
\end{theorem}

This theorem is proven in \S \ref{subsection:Lax_Painleve}.
The actions \eqref{eqns:intro_action_para} and \eqref{eqns:intro_compatibility_conditions} correspond to the $q$-Painlev\'e equations \eqref{eqns:qpainleve_intro} as explained in the following remark.
\begin{remark}\label{remark:relation_qps}
Equations \eqref{eqn:intro_T4action_f} and \eqref{eqn:intro_T3action_f} are equivalent to 
$q$-{\rm P}$_{\rm IV}$ \eqref{eqn:qp4_intro} and
$q$-${\rm P}_{\rm III}$ \eqref{eqn:qp3_intro}
by the following correspondences:
\begin{subequations}
\begin{align}
 &\bar{}=\IVT,\quad
 a=a_0,\quad
 b=a_1,\quad
 c=a_2,\quad
 t=\la,\quad
 f=f_0,\quad
 g=f_1,\quad
 h=f_2,\\
 &\bar{}=\IIIT,\quad
 a=\la^2,\quad
 b=a_2,\quad
 t=a_0,\quad
 f=f_0,\quad
 g=f_1,
\end{align}
\end{subequations}
respectively.
Moreover, letting 
\begin{equation}
 a_2=q^{1/2},
\end{equation}
and setting
\begin{equation}
 \tilde{}=\SIIIT,\quad
 t=a_0,\quad
 f=f_0,
\end{equation}
we obtain $q$-${\rm P}_{\rm II}$ \eqref{eqn:qp2_intro} from Equation \eqref{eqn:intro_TS3action_f}.
\end{remark}

\subsection{Background}\label{subsection:background}
Discrete Painlev\'e equations are nonlinear ordinary difference equations of second order, 
which include discrete analogues of the six Painlev\'e equations: P$_{\rm I}$, $\dots$, P$_{\rm VI}$. 
The geometric classification of discrete Painlev\'e equations,
based on types of rational surfaces connected to affine Weyl groups,
is well known\cite{SakaiH2001:MR1882403}.
Together with the Painlev\'e equations, 
they are now regarded as one of the most important classes of equations 
in the theory of integrable systems (see, e.g., \cite{GR2004:MR2087743}). 

In \cite{ABS2003:MR1962121,ABS2009:MR2503862,BollR2011:MR2846098,BollR2012:MR3010833,BollR:thesis}, 
Adler-Bobenko-Suris (ABS) and Boll classified polynomials $P$, say, of four variables into eleven types:
Q4, Q3, Q2, Q1, H3, H2, H1, D4, D3, D2, D1.
The first four types, the next three types and the last four types are collectively called $Q$-, $H^4$- and $H^6$-types, respectively.
The resulting polynomial $P$ satisfies the following properties.
\begin{description}
\item[(1) Linearity]
$P$ is linear in each argument, i.e., it has the following form:
\begin{equation}
 P(x_1,x_2,x_3,x_4)=A_1x_1x_2x_3x_4+\cdots+A_{16},
\end{equation}
where coefficients $A_i$ are complex parameters.
\item[(2) 3D consistency and tetrahedron property]
There exist a further seven polynomials of four variables: $P^{(i)}$, $i=1,\dots,7$,
which satisfy property {\bf(1)}
and a cube $C$ on whose six faces the following equations are assigned
\begin{subequations}
\begin{align}
 &P(x_0,x_1,x_2,x_{12})=0,
 &&P^{(1)}(x_0,x_2,x_3,x_{23})=0,\\
 &P^{(2)}(x_0,x_3,x_1,x_{31})=0,
 &&P^{(3)}(x_3,x_{31},x_{23},x_{123})=0,\\
 &P^{(4)}(x_1,x_{12},x_{31},x_{123})=0,
 &&P^{(5)}(x_2,x_{23},x_{12},x_{123})=0,
\end{align}
\end{subequations}
where the eight variables $x_0$, \dots, $x_{123}$ lie on the vertices of the cube,
in such a way that 
$x_{123}$ can be uniquely expressed in terms of the four variables $x_0$, $x_1$, $x_2$, $x_3$
({\it 3D consistency})
and moreover the following relations hold ({\it tetrahedron property}):
\begin{equation}
 P^{(6)}(x_0,x_{12},x_{23},x_{31})=0,\quad
 P^{(7)}(x_1,x_2,x_3,x_{123})=0.
\end{equation}
\end{description} 
Since these equations relate the vertices of the quadrilateral on a lattice,
they are often called quad-equations or lattice equations.

Some polynomials of ABS type are
\begin{align*}
 \text{Q1}&:Q1(x_1,x_2,x_3,x_4;\alpha_1,\alpha_2;\epsilon)\\
 &\quad=\alpha_1(x_1x_2+x_3x_4)-\alpha_2(x_1x_4+x_2x_3)
 -(\alpha_1-\alpha_2)(x_1x_3+x_2x_4)+\epsilon\alpha_1\alpha_2(\alpha_1-\alpha_2),\\
 \text{H3}&:H3(x_1,x_2,x_3,x_4;\alpha_1,\alpha_2;\delta;\epsilon)\\
 &\quad=\alpha_1(x_1x_2+x_3x_4)-\alpha_2(x_1x_4+x_2x_3)
 +({\alpha_1}^2-{\alpha_2}^2)\left(\delta+\cfrac{\epsilon}{\alpha_1\alpha_2}\,x_2x_4\right),\\
 \text{H1}&:H1(x_1,x_2,x_3,x_4;\alpha_1,\alpha_2;\epsilon)
 =(x_1-x_3)(x_2-x_4)+(\alpha_2-\alpha_1)(1-\epsilon x_2x_4),\\
 \text{D4}&:D4(x_1,x_2,x_3,x_4;\delta_1,\delta_2,\delta_3)
 =x_1x_3+x_2x_4+\delta_1x_1x_4+\delta_2x_3x_4+\delta_3,
\end{align*}
where $\alpha_1,\alpha_2\in\mathbb{C}^\ast$ and $\epsilon,\delta,\delta_1,\delta_2,\delta_3\in\{0,1\}$.
Many well known integrable \PDE s arise from assigning a polynomial of ABS type to quadrilaterals in the integer lattice $\bbZ^2$,
for example:
\begin{description}
\item[discrete Schwarzian KdV equation\cite{NC1995:MR1329559,NCWQ1984:MR763123}]
\begin{equation}\label{eqn:intro_DSKdV_1}
 Q1(U,\overline{U},\widehat{\overline{U}},\widehat{U};\alpha,\beta;0)=0
 ~\Leftrightarrow~
 \cfrac{(U-\overline{U})(\widehat{U}-\widehat{\overline{U}})}{(U-\widehat{U})(\overline{U}-\widehat{\overline{U}})}=\cfrac{\alpha}{\beta}\,;
\end{equation}
\item[lattice modified KdV equation\cite{NC1995:MR1329559,NQC1983:MR719638,ABS2003:MR1962121}]
\begin{equation}\label{eqn:intro_LMKdV_1}
 H3(U,\overline{U},-\widehat{\overline{U}},\widehat{U};\alpha,\beta;0;0)=0
 ~\Leftrightarrow~
 \cfrac{\widehat{\overline{U}}}{U}=\cfrac{\alpha\overline{U}-\beta\widehat{U}}{\alpha\widehat{U}-\beta\overline{U}}\,;
\end{equation}
\item[lattice potential KdV equation\cite{HirotaR1977:MR0460934,NC1995:MR1329559}]
\begin{equation}
 H1(U,\overline{U},\widehat{\overline{U}},\widehat{U};\alpha,\beta;0)=0
 ~\Leftrightarrow~
 (U-\widehat{\overline{U}})(\overline{U}-\widehat{U})=\alpha-\beta\,;
\end{equation}
\item[discrete version of Volterra-Kac-van Moerbeke equation\cite{NC1995:MR1329559}]
\begin{align}
 &D4(1-(\alpha^{-1}\beta-1)U,\widehat{U},\overline{U},-1+(\alpha^{-1}\beta-1)\widehat{\overline{U}};0,0,0)=0\notag\\
 &~\Leftrightarrow~
 \cfrac{\widehat{U}}{\overline{U}}=\cfrac{(\beta-\alpha)U-\alpha}{(\beta-\alpha)\widehat{\overline{U}}-\alpha}\,,
\end{align}
\end{description}
where 
\begin{equation}\label{eqn:intro_notation_1}
 U=U_{l,m},\quad
 \alpha=\alpha_l,\quad
 \beta=\beta_m,\quad
 \bar{ }:l\to l+1,\quad
 \hat{ }:m\to m+1,\quad
 l,m\in\mathbb{Z}.
\end{equation}
Throughout this paper, we refer to  such \PDE s as ABS equations.

We note that in general a hypercube is said to be multi-dimensionally consistent, if all cubes contained in the hypercube are 3D consistent (see property {\bf(2)} above).
Reductions of such ABS equations to ordinary difference equations have been found through several approaches \cite{JGTR2006:MR2271126,NP1991:MR1098879,GRSWC2005:MR2117991,FJN2008:MR2425981,HHJN2007:MR2303490,OrmerodCM2012:MR2997166,HHNS2015:MR3317164,OrmerodCM2014:MR3210633,AHJN2016:MR3509963}.
Our geometric-reduction method \cite{JNS2015:MR3403054,JNS2014:MR3291391,JNS:paper4,JNS:paper5} has shown how to obtain discrete Painlev\'e equations by studying geometric connections between these and ABS equations.

These equations are called integrable because they arise as compatibility conditions for associated linear problems called Lax pairs. 
The search for and construction of Lax pairs of discrete Painlev\'e equations has been a very active research area and the investigations have been carried out through many approaches. 
Noteworthy approaches include extensions of Birkhoff's study of linear $q$-difference equations\cite{JS1996:MR1403067,SakaiH2006:MR2266221,SakaiH2005:MR2177121},
periodic-type reductions from ABS equations or the discrete KP/UC hierarchy\cite{HayM2007:MR2371129,HHJN2007:MR2303490,OVHQ2014:MR3215839,OrmerodCM2012:MR2997166,OVQ2013:MR3030178,KNY2002:MR1958118,TsudaT2010:MR2563787,PNGR1992:MR1162062},
extensions of Schlesinger transformations
\cite{DT2014:arXiv1408.3778,DST2014:arXiv1302.2972,BoalchP2009:MR2500553},
search for linearizable curves in initial-value space \cite{YamadaY2009:MR2506170,YamadaY2011:MR2836394,KNY2015:arXiv150908186K},
Pad\'e approximation or interpolation\cite{IkawaY2013:MR3061504,NagaoH2015:MR3323664,NTY2013:MR3077695} 
and the theory of orthogonal polynomials\cite{WO2012:MR3007262,WitteNS2015:MR3413576,OWF2011:MR2819929,BianeP2014:MR3221944,BA2010:MR2578525,BB2003:MR1962463}.
 
However, the construction of Lax pairs for each case in the literature has been carried out in different ways for different equations on the same surface. 
In contrast, in the case we study, $q$-P$_{\rm IV}$ \eqref{eqn:qp4_intro}, $q$-P$_{\rm III}$ \eqref{eqn:qp3_intro} and $q$-P$_{\rm II}$ \eqref{eqn:qp2_intro}
are all obtained on the $A_5^{(1)}$-surface. 
\subsection{Plan of the paper}
The plan of this paper is as follows.
In \S \ref{section:lax_qPs_A5}, we construct the Lax pairs of the \PDE s on the 4-dimensional integer lattice.
Then, we obtain the Lax pairs of the $A_5^{(1)}$-surface $q$-Painlev\'e equations from them through geometric reduction.
Some concluding remarks are given in \S \ref{ConcludingRemarks}.
\section{The Lax pairs of the $A_5^{(1)}$-surface $q$-Painlev\'e equations}
\label{section:lax_qPs_A5}
\subsection{The \PDE s on the lattice $\bbZ^4$}
In this section, we consider the \PDE s on the 4-dimensional integer lattice $\bbZ^4$.

In the same way that the lattice $\bbZ^2$ can be constructed by tiling the plane with squares, we construct the lattice $\bbZ^4$ by tiling it with 4-dimensional hypercubes (i.e. 4-cubes).
We obtain \PDE s on the lattice $\bbZ^4$ in a similar manner to the constructions of the ABS equations (see \S \ref{subsection:background}).
Indeed, assigning the function $u$ and quad-equations of ABS type to the vertices and faces of each 4-cube,
we obtain the following system of ABS equations:
\begin{subequations}\label{eqns:4D_PDEs_u}
\begin{align}
 &\cfrac{u(\bml+\ep_1+\ep_2)}{u(\bml)}
 =\cfrac{u(\bml+\ep_1)- \cfrac{\be_{l_2}}{\al_{l_1}}\,u(\bml+\ep_2)}{\cfrac{\be_{l_2}}{\al_{l_1}}\,u(\bml+\ep_1)-u(\bml+\ep_2)},\quad
 \cfrac{u(\bml+\ep_1+\ep_4)}{u(\bml)}+\cfrac{u(\bml+\ep_4)}{u(\bml+\ep_1)}=-\al_{l_1} K_{l_4},
 \label{eqns:4D_PDEs_u_1}\\
 &\cfrac{u(\bml+\ep_2+\ep_3)}{u(\bml)}
 =\cfrac{u(\bml+\ep_2)-\cfrac{\ga_{l_3}}{\be_{l_2}}\,u(\bml+\ep_3)}{\cfrac{\ga_{l_3}}{\be_{l_2}}\,u(\bml+\ep_2)-u(\bml+\ep_3)},\quad
  \cfrac{u(\bml+\ep_2+\ep_4)}{u(\bml)}+\cfrac{u(\bml+\ep_4)}{u(\bml+\ep_2)}=-\be_{l_2} K_{l_4},\\ 
 &\cfrac{u(\bml+\ep_3+\ep_1)}{u(\bml)}
 =\cfrac{u(\bml+\ep_3)-\cfrac{\al_{l_1}}{\ga_{l_3}}\,u(\bml+\ep_1)}{\cfrac{\al_{l_1}}{\ga_{l_3}}\,u(\bml+\ep_3)- u(\bml+\ep_1)},\quad
 \cfrac{u(\bml+\ep_3+\ep_4)}{u(\bml)}+\cfrac{u(\bml+\ep_4)}{u(\bml+\ep_3)}
 =-\ga_{l_3} K_{l_4},
\end{align}
\end{subequations}
where $\bml=\sum_{i=1}^4 l_i \ep_i\in\bbZ^4$.
Here, $u(\bml)$ is the function on the lattice $\bbZ^4$
and $\Set{\al_l}{l\in\bbZ}$, $\Set{\be_l}{l\in\bbZ}$, $\Set{\ga_l}{l\in\bbZ}$ and $\Set{K_l}{l\in\bbZ}$ are complex parameters.
We note that each left hand sides of the equations \eqref{eqns:4D_PDEs_u} is called H3$_{\de=0}$ (or H3$_{\de=0}^{\epsilon=0}$) 
and each right hand sides of them is called D4 in the ABS classification\cite{ABS2003:MR1962121,ABS2009:MR2503862,BollR2011:MR2846098}.

\begin{figure}[t]
\includegraphics[width=0.6\textwidth]{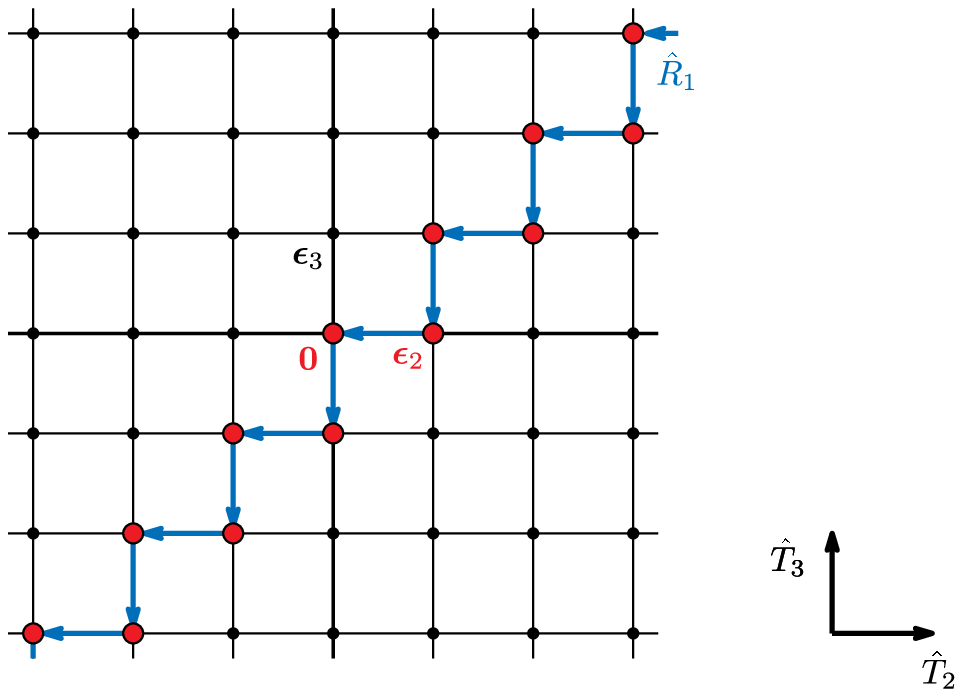}
\caption{Moving of the mapping $\hR_1$ around the origin $\bm{0}\in\bbZ^4$.}
\label{fig:moving_hR1}
\end{figure}

In the lattice $\bbZ^4$, there are four orthogonal directions, which naturally give rise to four translation operators.
In our case, these result in actions on the variable $u(\bml)$ and the parameters $\al_l$, $\be_l$, $\ga_l$, $K_l$
and lead to the transformations $\hT_i$, $i=1,\dots,4$, by the following actions:
\begin{subequations}
\begin{align}
 &\hT_1:(u(\bml),\al_l,\be_l,\ga_l,K_l)\mapsto(u(\bml+\ep_1),\al_{l+1},\be_l,\ga_l,K_l),\\
 &\hT_2:(u(\bml),\al_l,\be_l,\ga_l,K_l)\mapsto(u(\bml+\ep_2),\al_l,\be_{l+1},\ga_l,K_l),\\
 &\hT_3:(u(\bml),\al_l,\be_l,\ga_l,K_l)\mapsto(u(\bml+\ep_3),\al_l,\be_l,\ga_{l+1},K_l),\\
 &\hT_4:(u(\bml),\al_l,\be_l,\ga_l,K_l)\mapsto(u(\bml+\ep_4),\al_l,\be_l,\ga_l,K_{l+1}).
\end{align}
\end{subequations}
In the two-dimensional slice given by $\hT_2$ and $\hT_3$, we define the diagonal region
 $\calR=\calr^{(1)}\cup\calr^{(2)}\subset\bbZ^4$ where
\begin{equation}
 \calr^{(1)}=\set{\sum_{i=1}^4 l_i \ep_i}{l_i\in\bbZ,~ l_3=l_2-1},\quad
 \calr^{(2)}=\set{\sum_{i=1}^4 l_i \ep_i}{l_i\in\bbZ,~ l_3=l_2}.
\end{equation}
We also define the action of a staircase path $\hR_1$ (see Figure \ref{fig:moving_hR1})
for one of the discrete Painlev\'e equations considered below:
\begin{equation}
 \hR_1(\bml)
 =\begin{cases}
 \bml-\ep_2&\text{if}\quad \bml\in\calr^{(1)},\\
 \bml-\ep_3&\text{if}\quad \bml\in\calr^{(2)},
 \end{cases}
\end{equation}
on the variable $u(\bml)$, $\bml\in\calR$, and the parameters $\al_l$, $\be_l$, $\ga_l$, $K_l$ by
\begin{equation}
  \hR_1:\left(u(\bml),\al_l,\be_l,\ga_l,K_l\right)\mapsto \left(u(\hR_1(\bml)),\al_l,\ga_{l-1},\be_l,K_l\right).
\end{equation}

We define the actions of $\hT_i$, $i=1,\dots,4$, and $\hR_1$ on the infinite extension field of the complex field $\bbC$,
generated by $\Set{u(\bml)}{\bml\in\bbZ^4\text{ or }\calR}$, $\Set{\al_l}{l\in\bbZ}$, $\Set{\be_l}{l\in\bbZ}$, 
$\Set{\ga_l}{l\in\bbZ}$ and $\Set{K_l}{l\in\bbZ}$, as automorphisms.
Henceforth, if a new quantity $x$ is added, 
we extend the field on which $\hT_i$, $i=1,\dots,4$, and $\hR_1$ act as the automorphisms by adding the generator $x$.
Note that throughout this paper every field is of characteristic zero.
Moreover, when the field is generated by $\Set{x_1,\dots,x_k}{}$,
a mapping $w\in\langle \hT_1,\dots,\hT_4,\hR_1\rangle$ acts on an arbitrary function $F=F(x_1,\dots,x_k)$ by the following:
\begin{equation}
 w(F)=F(w(x_1),\dots,w(x_k)).
\end{equation}
For convenience, throughout this paper we use the following notation for the combined transformation of arbitrary mappings $w$ and $w'$:
\begin{equation}
 ww':=w\circ w'.
\end{equation}

\subsection{Lax pair of the system of the \PDE s \eqref{eqns:4D_PDEs_u}}
\label{subsection:Lax_PDEs}
In this section, we construct the Lax pair of the system of the \PDE s \eqref{eqns:4D_PDEs_u}
following the method given in \cite{BS2002:MR1890049,NijhoffFW2002:MR1912127,WalkerAJ:thesis}.

The key to the construction of the Lax pair of the system of the \PDE s \eqref{eqns:4D_PDEs_u} is
to introduce fifth direction: $\bbZ^4\ni \bml\mapsto \bml+\ep_5\in\bbZ^5$ 
arising from the extension of the multi-dimensionally consistent 4-cube to the multi-dimensionally consistent 5-cube
and to regard it as a virtual direction.
As a result, we obtain the following additional relations:
\begin{subequations}\label{eqns:PDEs_baru}
\begin{align}
 &\cfrac{\bar{u}(\bml+\ep_1)}{u(\bml)}
 =-\cfrac{\al_{l_1} u(\bml+\ep_1)-\mu \bar{u}(\bml)}{\al_{l_1} \bar{u}(\bml)-\mu u(\bml+\ep_1)},\quad
 \cfrac{\bar{u}(\bml+\ep_2)}{u(\bml)}
 =-\cfrac{\be_{l_2} u(\bml+\ep_2)-\mu \bar{u}(\bml)}{\be_{l_2} \bar{u}(\bml)-\mu u(\bml+\ep_2)},
 \label{eqns:baru_1_2}\\
 &\cfrac{\bar{u}(\bml+\ep_3)}{u(\bml)}
 =-\cfrac{\ga_{l_3} u(\bml+\ep_3)-\mu \bar{u}(\bml)}{\ga_{l_3} \bar{u}(\bml)-\mu u(\bml+\ep_3)},\quad
 \cfrac{\bar{u}(\bml+\ep_4)}{u(\bml)}+\cfrac{u(\bml+\ep_4)}{\bar{u}(\bml)}=-\mu K_{l_4},
\end{align}
\end{subequations}
where $\bar{u}(\bml)=u(\bml+\ep_5)$, $\bml=\sum_{i=1}^4 l_i \ep_i\in\bbZ^4$ and $\mu$ is the additional complex parameter.
We distinguish the function $\bar{u}(\bml)$ from $u(\bml)$.
Then, each of Equations \eqref{eqns:PDEs_baru} can be regarded as the first order discrete system of Riccati type of the quantity $\bar{u}(\bml)$,
which is linearizable.
Indeed, substituting 
\begin{equation}\label{eqn:u_FG}
 \bar{u}(\bml)=\cfrac{F(\bml)}{G(\bml)},
\end{equation}
in Equations \eqref{eqns:PDEs_baru}
and dividing them into the numerators and the denominators with the vector $\Psi=\Psi(\bml)$ defined by
\begin{equation}
 \Psi=\begin{pmatrix}F(\bml)\\G(\bml)\end{pmatrix},
\end{equation}
we obtain the following linear systems:
\begin{subequations}\label{eqns:lax_PDEs_1}
\begin{align}
 &\hT_1(\Psi)
 =\de^{(1)}
 \begin{pmatrix}
 \cfrac{\mu}{\al_{l_1}}&-\hT_1(u(\bml))\\
 \cfrac{1}{u(\bml)}&-\cfrac{\mu}{\al_{l_1}}\,\cfrac{\hT_1(u(\bml))}{u(\bml)}
 \end{pmatrix}
 \Psi,\quad
 \hT_2(\Psi)
 =\de^{(2)}
 \begin{pmatrix}
 \cfrac{\mu}{\be_{l_2}}&-\hT_2(u(\bml))\\
 \cfrac{1}{u(\bml)}&-\cfrac{\mu}{\be_{l_2}}\,\cfrac{\hT_2(u(\bml))}{u(\bml)}
 \end{pmatrix}
 \Psi,\label{eqns:psi_T1_T2}\\
 &\hT_3(\Psi)
 =\de^{(3)}
 \begin{pmatrix}
 \cfrac{\mu}{\ga_{l_3}}&-\hT_3(u(\bml))\\
 \cfrac{1}{u(\bml)}&-\cfrac{\mu}{\ga_{l_3}}\,\cfrac{\hT_3(u(\bml))}{u(\bml)}
 \end{pmatrix}
 \Psi,\quad
 \hT_4(\Psi)
 =\de^{(4)}
 \begin{pmatrix}
 -\mu K_{l_4}&-\hT_4(u(\bml))\\
 \cfrac{1}{u(\bml)}&0
 \end{pmatrix}
 \Psi,
\end{align}
\end{subequations} 
where $\bml=\sum_{i=1}^4 l_i \ep_i$ and $\de^{(i)}=\de^{(i)}(\bml)$, $i=1,\dots,4$, are arbitrary decoupling factors.
The linear systems \eqref{eqns:lax_PDEs_1} are the Lax pair of the system of the \PDE s \eqref{eqns:4D_PDEs_u},
that is, the compatibility condition of each pair of the linear systems \eqref{eqns:lax_PDEs_1} 
gives one of Equations \eqref{eqns:4D_PDEs_u}.
For example, the compatibility condition of $\hT_1$ and $\hT_2$:
\begin{equation}
 \hT_1\hT_2(\Psi)=\hT_2\hT_1(\Psi),
\end{equation}
gives the left hand side of Equation \eqref{eqns:4D_PDEs_u_1} and 
\begin{equation}
  \hT_1(\de^{(2)})\de^{(1)}=\hT_2(\de^{(1)})\de^{(2)}.
\end{equation}
Note that we define the action of a transformation $w\in\langle \hT_1,\dots,T_4,\hR_1\rangle$ on the vector $\Psi$ by
\begin{equation}
  w(\Psi(\bml))=\Psi(w(\bml)).
\end{equation}
\subsection{Geometric reduction of the \PDE s \eqref{eqns:4D_PDEs_u}}
\label{subsection:geometric_reduction}
In this section, we apply the geometric reduction considered in \cite{JNS2014:MR3291391}
to the \PDE s \eqref{eqns:4D_PDEs_u}.
Moreover, we obtain the $q$-Painlev\'e equations \eqref{eqns:intro_compatibility_conditions}
from the reduced \PDE s \eqref{eqns:omega_fundamental}. 

Let
\begin{align}
 &h_{l_1,l_2,l_3,l_4}=\ii^{\,\log(q^{l_1+l_2+l_3+l_4}\hal\hbe\hga\la)/\log{q}}(q^{l_4}\la)^{\log(q^{l_2}\hbe)/\log{q}},\\
 &a_0=q\cfrac{\hal}{\hga},\quad
 a_1=\cfrac{\hbe}{\hal},\quad
 a_2=\cfrac{\hga}{\hbe},
\end{align}
where $a_0a_1a_2=q$, $\ii=\sqrt{-1}$ and $\hal,\hbe,\hga,\la,q\in\bbC$ are parameters.
By letting
\begin{equation}
 u(\bml)=h_{l_1,l_2,l_3,l_4}\,\omega(\bml),
\end{equation}
where $\bml=\sum_{i=1}^4l_i\ep_i$,
and imposing the following periodic condition for $\bml\in\bbZ^4$:
\begin{equation}\label{eqn:111periodic}
 \omega(\bml+\ep_1+\ep_2+\ep_3)=\omega(\bml),
\end{equation}
with the following condition of the parameters
$\Set{\al_l}{l\in\bbZ}$, $\Set{\be_l}{l\in\bbZ}$, $\Set{\ga_l}{l\in\bbZ}$ and $\Set{K_l}{l\in\bbZ}$:
\begin{equation}\label{eqn:111_4d_para}
 \al_l=q^l\hal,\quad
 \be_l=q^l\hbe,\quad
 \ga_l=q^l\hga,\quad
 K_l=\cfrac{q^{2l+1}\la^2-1}{q^l\la},
\end{equation}
the \PDE s \eqref{eqns:4D_PDEs_u} are reduced to the following \PDE s:
{\allowdisplaybreaks
\begin{subequations}\label{eqns:omega_fundamental}
\begin{align}
 &\cfrac{\omega(\bml+\ep_1+\ep_2)}{\omega(\bml)}
 =\cfrac{\omega(\bml+\ep_1)-q^{-l_1+l_2+l_4}\la a_1 \omega(\bml+\ep_2)}
 {q^{l_4}\la\left(q^{l_4}\la\,\omega(\bml+\ep_2)-q^{-l_1+l_2}a_1\omega(\bml+\ep_1)\right)},\\
 &\cfrac{\omega(\bml+\ep_2+\ep_3)}{\omega(\bml)}
 =\cfrac{q^{l_4}\la\, \omega(\bml+\ep_2)-q^{-l_2+l_3}a_2 \omega(\bml+\ep_3)}
 {q^{l_4}\la\left(\omega(\bml+\ep_3)-q^{-l_2+l_3+l_4}\la a_2 \omega(\bml+\ep_2)\right)},\\
 &\cfrac{\omega(\bml+\ep_3+\ep_1)}{\omega(\bml)}
 =\cfrac{\omega(\bml+\ep_3)-q^{l_1-l_3-1}a_0\omega(\bml+\ep_1)}
 {\omega(\bml+\ep_1)-q^{l_1-l_3-1}a_0\omega(\bml+\ep_3)},\\
 &\cfrac{\omega(\bml+\ep_1+\ep_4)}{\omega(\bml)}
 -\cfrac{\omega(\bml+\ep_4)}{\omega(\bml+\ep_1)}
 =\cfrac{q^{2l_4+1}\la^2-1}{q^{-l_1+l_2+l_4}a_1\la},\\
 &\cfrac{\omega(\bml+\ep_2+\ep_4)}{\omega(\bml)}
 -\cfrac{1}{q^{2l_4+1}\la^2}\,\cfrac{\omega(\bml+\ep_4)}{\omega(\bml+\ep_2)}
 =\cfrac{q^{2l_4+1}\la^2-1}{q^{2l_4+1}\la^2},\\
 &\cfrac{\omega(\bml+\ep_3+\ep_4)}{\omega(\bml)}
 -\cfrac{\omega(\bml+\ep_4)}{\omega(\bml+\ep_3)}
 =\cfrac{q^{-l_2+l_3}a_2(q^{2l_4+1}\la^2-1)}{q^{l_4}\la}.
\end{align}
\end{subequations}
}
Then, the actions of $\hT_i$, $i=1,\dots,4$, and $\hR_1$ on the $\omega$-function and 
the parameters $a_i$, $i=0,1,2$, $\la$ and $q$ are given by
\begin{subequations}
\begin{align}
 &\hT_1:(\omega(\bml),a_0,a_1,a_2,\la,q)\mapsto(\omega(\bml+\ep_1),qa_0,q^{-1}a_1,a_2,\la,q),\\
 &\hT_2:(\omega(\bml),a_0,a_1,a_2,\la,q)\mapsto(\omega(\bml+\ep_2),a_0,qa_1,q^{-1}a_2,\la,q),\\
 &\hT_3:(\omega(\bml),a_0,a_1,a_2,\la,q)\mapsto(\omega(\bml+\ep_3),q^{-1}a_0,a_1,qa_2,\la,q),\\
 &\hT_4:(\omega(\bml),a_0,a_1,a_2,\la,q)\mapsto(\omega(\bml+\ep_4),a_0,a_1,a_2,q\la,q),\\
 &\hR_1:(\omega(\bml),a_0,a_1,a_2,\la,q)\mapsto(\omega(\hR_1(\bml)),a_0a_2,q^{-1}a_1a_2,q{a_2}^{-1},\la,q).
\end{align}
\end{subequations}

\begin{figure}[t]
\includegraphics[width=0.8\textwidth]{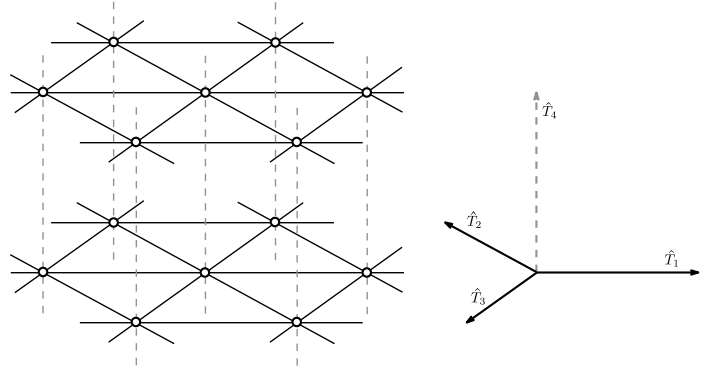}
\caption{$(A_2+A_1)^{(1)}$-lattice}
\label{fig:A2A1_lattice}
\end{figure}

\begin{remark}
If we consider the actions of the transformations $\hT_i$, $i=1,\dots,4$,
only on the $\omega$-function and the parameters $a_i$, $i=0,1,2$, $\la$ and $q$,
then the transformation $\hT_1\hT_2\hT_3$ can be regarded as the identity mapping:
\begin{equation}
 \hT_1\hT_2\hT_3=\mathrm{Id}.
\end{equation}
In this situation, we can proceed as if the reduction acts on the lattice $\bbZ^4$, 
that is, the lattice $\bbZ^4$ is reduced to the $(A_2+A_1)^{(1)}$-lattice (see Figure \ref{fig:A2A1_lattice}):
\begin{equation}
 \bbZ^4 \to \bbZ^4/\bbZ(\ep_1+\ep_2+\ep_3).
\end{equation}
The reduction from the lattice $\bbZ^4$  to the $(A_2+A_1)^{(1)}$-lattice with the \PDE s
is referred to as the {\it geometric reduction}\cite{JNS2014:MR3291391}.
\end{remark}

The relations \eqref{eqns:omega_fundamental} are equivalent to the essential relations of 
the $\omega$-lattice of type $A_5^{(1)}$ constructed in \cite{JNS2015:MR3403054}
with the following correspondences:
\begin{equation}
 \omega(\bm{0})=\omega_0,\quad
 \hT_i=T_i,~i=1,\dots,4,\quad
 \hR_1=R_1.
\end{equation}
Therefore, considering the \PDE s \eqref{eqns:omega_fundamental}
is equivalent to considering the $\omega$-lattice.
In this case, the $\omega$-lattice is said to have a {\it reduced hypercube structure}.
This reduced hypercube structure turns out to be essential in the construction of Lax pairs 
for discrete Painlev\'e equations.
Since the $q$-Painlev\'e equations \eqref{eqns:intro_compatibility_conditions}
are discrete dynamical systems on the $\omega$-lattice of type $A_5^{(1)}$ as shown in \cite{JNS2015:MR3403054},
we can obtain the $q$-Painlev\'e equations \eqref{eqns:intro_compatibility_conditions} 
from the $\omega$-function with the essential relations \eqref{eqns:omega_fundamental} as follows.
By letting
\begin{equation}\label{eqn:define_f_section4}
 f_0=\cfrac{\omega(\ep_1)}{\omega(\ep_1+\ep_2)}\,,\quad
 f_1=\la \cfrac{\omega(\ep_1+\ep_2)}{\omega(\bm{0})}\,,\quad
 f_2=\la \cfrac{\omega(\bm{0})}{\omega(\ep_1)}\,,
\end{equation}
where $f_0f_1f_2=\la^2$,
we obtain the $q$-Painlev\'e equations \eqref{eqns:intro_compatibility_conditions} with the following correspondences:
\begin{equation}
 \IVT=\hT_4,\quad \IIIT=\hT_3^{~-1}\hT_2^{~-1},\quad \SIIIT=\hR_1.
\end{equation}

\subsection{Lax pairs of the $A_5^{(1)}$-surface $q$-Painlev\'e equations}
\label{subsection:Lax_Painleve}
In this section, using the linear systems \eqref{eqns:lax_PDEs_1} and the connection between
the \PDE s \eqref{eqns:4D_PDEs_u} and the $A_5^{(1)}$-surface $q$-Painlev\'e equations via the geometric reduction,
we construct the Lax pairs of the $q$-Painlev\'e equations \eqref{eqns:intro_compatibility_conditions}.

We first define the vector $\phi$ by
\begin{equation}
 \Psi(\bm{0})=\begin{pmatrix}h_{0,0,0,0}\,\omega(\bm{0})&0\\0&1\end{pmatrix}\phi.
\end{equation}
Then, from the linear systems \eqref{eqns:lax_PDEs_1}, we obtain the following linear systems:
\begin{subequations}
\begin{align}
 &\hT_1(\phi)
 =\de^{(1)}
 \begin{pmatrix}
 -\cfrac{\ii f_2}{\la}\,x&-1\\
 1&-\cfrac{\ii \la}{f_2}\,x
 \end{pmatrix}
 \phi,\\
 &\hT_2(\phi)
 =\de^{(2)}
 \begin{pmatrix}
 - \cfrac{\ii{T_1}^{-1}(f_0)}{a_1\la}\,x&-1\\
 1&- \cfrac{\ii\la}{a_1 {T_1}^{-1}(f_0)}\,x
 \end{pmatrix}
 \phi,\\
 &\hT_3(\phi)
 =\de^{(3)}
 \begin{pmatrix}
 - \cfrac{\ii a_0 T_3(f_1)}{q\la}\,x&-1\\
 1&- \cfrac{\ii a_0\la}{q T_3(f_1)}\,x
 \end{pmatrix}
 \phi,\\
 &\hT_4(\phi)
 =\de^{(4)}
 \begin{pmatrix}
 \cfrac{\ii(q\la^2-1) f_2}
 {\la\left(1+a_1(1+a_2 f_2)f_1\right)}\,x&-1\\
 1&0
 \end{pmatrix}
 \phi,\\
 &\hR_1(\phi)=\hT_3^{~-1}(\phi),
\end{align}
\end{subequations}
where 
\begin{equation}
 x=\cfrac{\mu}{\hal}.
\end{equation}
Next, let us define the transformations $\spT$, $\IVT$, $\IIIT$ and $\SIIIT$ by
\begin{equation}
 \spT=\hT_3^{~-1}\hT_2^{~-1}\hT_1^{~-1},\quad
 \IVT=\hT_4,\quad
 \IIIT=\hT_3^{~-1}\hT_2^{~-1},\quad
 \SIIIT=\hR_1.
\end{equation}
The actions of $\spT$, $\IVT$, $\IIIT$ and $\SIIIT$ on the spectral parameter $x$ are given by
\begin{equation}
 \spT(x)=qx,\quad
 \IVT(x)=\IIIT(x)=\SIIIT(x)=x,
\end{equation}
while those on the wave function $\phi$ are given by the following:
\begin{subequations}
\begin{align}
 &\spT(\phi)=\cfrac{\de^{\rm (SP)}}{(1-q^2 x^2)(1-{a_0}^2{a_2}^2 x^2)(1-{a_0}^2 x^2)}\,A\,\phi,\quad
 \IVT(\phi)=\de^{\rm (IV)}\,B_{\rm IV}\,\phi,\\
 &\IIIT(\phi)=\cfrac{\de^{\rm (III)}}{(1-{a_0}^2{a_2}^2 x^2)(1-{a_0}^2 x^2)}\,B_{\rm III}\,\phi,\quad
 \SIIIT(\phi)=\cfrac{\de^{\rm (SIII)}}{1-{a_0}^2 x^2}\,B_{\rm SIII}\,\phi,
\end{align}
\end{subequations}
where
\begin{subequations}
\begin{align}
 &\de^{\rm (SP)}=\cfrac{1}{{\hT_1}^{~-1}{\hT_2}^{~-1}{\hT_3}^{~-1}(\de^{(1)}){\hT_2}^{~-1}{\hT_3}^{~-1}(\de^{(2)}){\hT_3}^{~-1}(\de^{(3)})},\quad
 \de^{\rm (IV)}=\de^{(4)},\\
 &\de^{\rm (III)}=\cfrac{1}{{\hT_2}^{~-1}{\hT_3}^{~-1}(\de^{(2)}){\hT_3}^{~-1}(\de^{(3)})},\quad
 \de^{\rm (SIII)}=\cfrac{1}{{\hT_3}^{~-1}(\de^{(3)})}.
\end{align}
\end{subequations}
Here,  the $2\times2$ matrices $A$, $B_{\rm IV}$, $B_{\rm III}$ and $B_{\rm SIII}$ are given by 
Equations \eqref{eqn:linear_martrix_A} and \eqref{eqns:linear_martrix_B}.
Therefore, we finally obtain Theorem \ref{maintheorem_Lax} by the following correspondences:
\begin{equation}
 \de^{(1)}=\cfrac{1}{1-x^2}\,,\quad
 \de^{(2)}=\cfrac{1}{1-q^{-2}{a_0}^2{a_2}^2x^2}\,,\quad
 \de^{(3)}=\cfrac{1}{1-q^{-2}{a_0}^2x^2}\,,\quad
 \de^{(4)}=1,
\end{equation}
which give
\begin{subequations}
\begin{align}
 &\de^{\rm (SP)}=(1-q^2 x^2)(1-{a_0}^2{a_2}^2 x^2)(1-{a_0}^2 x^2),\quad
 \de^{\rm (IV)}=1,\\
 &\de^{\rm (III)}=(1-{a_0}^2{a_2}^2 x^2)(1-{a_0}^2 x^2),\quad
 \de^{\rm (SIII)}=1-{a_0}^2 x^2.
\end{align}
\end{subequations}

\section{Concluding remarks}
\label{ConcludingRemarks}
In this paper, we provided a comprehensive method for constructing Lax pairs of discrete Painlev\'e equations
by using a reduced hypercube structure.
As an example, we constructed the Lax pairs of the $q$-Painlev\'e equations \eqref{eqns:qpainleve_intro}.
As remarked earlier, the discrete Painlev\'e equations studied in this paper all share one $A$-matrix in the Lax pairs.

It is possible that we could broaden the action of the affine Weyl group of type $(A_2+A_1)^{(1)}$, 
which is the symmetry group for the $A_5^{(1)}$-surface $q$-Painlev\'e equations,
by adding the actions on the wave function $\phi$ and the spectral parameter $x$.
This implies that we can construct Lax pairs of ``all" $A_5^{(1)}$-surface $q$-Painlev\'e equations,
which share one spectral linear problem.
We will discuss this possibility in a forthcoming paper.
Another interesting future direction of research is to extend our method to other surface types of discrete Painlev\'e equations.
\subsection*{Acknowledgment}
The authors would like to express their sincere thanks to Prof. Y. Yamada and Dr. S. Lobb for inspiring and fruitful discussions.
This research was supported by an Australian Laureate Fellowship \# FL120100094 and grants \# DP130100967 and \# DP160101728 from the Australian Research Council.

\def\cprime{$'$} \def\cprime{$'$}

\end{document}